\documentstyle[aps,prl,twocolumn,epsf]{revtex}
\begin{document}
\draft
\title{Decay-Produced Neutrino Hot Dark Matter}
\author{Steen Hannestad}
\address{Theoretical Astrophysics Center, Institute of Physics and Astronomy,
University of Aarhus, 
DK-8000 \AA rhus C, Denmark}
\date{\today}
\maketitle

\begin{abstract}
We investigate different types of neutrino hot dark matter
with respect
to structure formation and anisotropies in the 
cosmic microwave background radiation (CMBR). 
The possibility of neutrino hot dark
matter produced through the decay of a heavier neutrino by the
process $\nu_H \to \nu_L + \phi$, where $\phi$ is a scalar particle,
is discussed in detail. This type of dark matter can possibly
be distinguished observationally from the standard neutrino
dark matter by using new
CMBR data from the upcoming satellite missions MAP and PLANCK.

\end{abstract}


The cold dark matter (CDM) model \cite{ostriker}
forms the basis for all the currently
viable models of the way structures form in our universe. In its simplest
form it consists of a critical ($\Omega=1$) density universe whose 
energy density is almost solely ($\Omega_{\rm CDM} \simeq 0.95$)
in the form of non-relativistic 
non-interacting particles. In addition there is a small fraction of
baryons ($\Omega_B \simeq 0.05$). Structures in this model grow from
the original quantum fluctuations produced during the inflation era.
The trouble with the simple CDM model is that, if the density fluctuation
power spectrum is normalised to the CMBR quadrupole
anisotropy measured by the COBE satellite, there are far too large
fluctuations on small scales \cite{ostriker}.
Any realistic model must therefore include something that dampens
small scale fluctuations. Several possible solutions to this problem
have been proposed, for example the inclusion of vacuum energy as a
dominant component or tilting the initial power spectrum from inflation
\cite{SP96}. Perhaps the most promising model is the so-called 
cold+hot dark matter scenario (CHDM), where a component of 
hot dark matter (HDM) is included; particles that only become
non-relativistic around the epoch of matter-radiation equality. 
Such light particles
free stream and thereby smear out fluctuations below a certain length
scale. This produces the desired result, namely that fluctuations on
small scales should be lessened relative to the CDM model. A good fit
to observational data can be obtained with $\Omega_{\text{HDM}} 
\simeq 0.25$
in a flat ($\Omega = 1$) universe.

A light (eV) neutrino species seems by far the most realistic candidate
for hot dark matter. 
If only one standard neutrino species is available, its mass
should be roughly 5-7 eV if $\Omega_{\rm HDM} \simeq 0.25$ and 
$h_0 \simeq 0.5$ ($h_0$ is the Hubble parameter in units of 
100 km s$^{-1}$ Mpc$^{-1}$). 
However, 
numerical simulations have shown that a better fit can be obtained if
there are two neutrino species available with half that mass
(this model is referred to as 2$\nu$CHDM); that is,
a mass of roughly 2-3 eV \cite{PHKC95}. 
This would demand a degenerate mass hierarchy
or, alternatively, a four-component eV neutrino such as a light Dirac
neutrino. Unfortunately, the sterile components of a light Dirac neutrino 
do not come into thermal equilibrium after the QCD phase transition
and therefore they only contribute very little to the cosmic energy
density.
A more plausible method of producing neutrino dark matter is through
the decay of a heavy neutrino to a final state containing light neutrinos,
mimicking a degenerate
mass hierarchy. 

There are several possible decay channels for a heavy 
neutrino to produce light neutrinos, for instance the radiative decay
$\nu_H \to \nu_L \gamma$, the flavour violating weak process
$\nu_H \to \nu_L \bar{\nu}_L \nu_L$ or a decay of the type 
$\nu_H \to \nu_L \phi$, where $\phi$ is a scalar or pseudo-scalar particle
like the majoron \cite{MPbook}. 
Decays containing electromagnetically interacting
particles are strongly constrained by observations so that in essence
only decay to an ``invisible'' final state is allowed
\cite{raffelt}. We shall look
at the simplest example of such a decay, namely the $\nu_H \to \nu_L \phi$
process. Specifically we assume $\nu_H=\nu_\tau$ and $\nu_L=\nu_\mu$.
$\nu_e$ is assumed to be massless. 
A decay of this type occurs naturally in the so-called majoron models.
In these models, neutrinos acquire mass by interaction
with a new Higgs field and the spontaneous breakdown of the global 
symmetry results in a Nambu-Goldstone boson called the majoron
\cite{MPbook}.
In some of these models neutrinos have large Yukawa-type couplings 
to the majoron so that fast decays of the type $\nu_H \to \nu_L \phi$
are permitted. The coupling strength is very difficult to constrain
experimentally,
although a sufficiently large coupling would result in neutrinoless
double-beta decays with too short lifetimes \cite{MPbook}.
For masses and lifetimes in the range we are interested in there are
currently no direct experimental constraints.

Neutrinos are in thermodynamic equilibrium with the cosmic plasma
until the universe reaches a temperature of roughly 1 MeV, so that if the
decay takes place prior to this freeze-out temperature, the
additional density of light neutrinos will be washed out because of
thermal equilibration.
In the following we shall therefore always 
assume that the decay takes place subsequent to neutrino decoupling,
but prior to matter-radiation equality. In this case the number density
of light neutrinos is twice that of a standard decoupled neutrino
species, as in the 2$\nu$CHDM model.
Note that if the decay takes place after decoupling, but at a temperature
higher than $T \simeq 0.01$ MeV, the synthesis of light elements may
be changed significantly, excluding some regions of mass-lifetime
parameter space \cite{hannestad}.

The purpose of this paper is to investigate observational features of this 
``decaying neutrino cold+hot dark matter'' (D$\nu$CHDM) scenario
and compare it with the standard 2$\nu$CHDM model. 
There are several differences between this decay
scenario and the 2$\nu$CHDM model. First of all, the distribution
of light neutrinos will in general be non-thermal so that, even though
the number density of neutrinos in the two scenarios is equal, the
energy density is different. Second, the amount of relativistic energy
density present will be different in the two cases. 
We will discuss the possibility that new CMBR
experiments will allow us to distinguish between these different models of
hot dark matter.
In order to calculate CMBR and matter power spectra we have used the
CMBFAST program developed by Seljak and Zaldarriaga \cite{SZ96}.

The dynamics of the decay can be quite simply described
with only two parameters, the mass, $m_H$, and lifetime, $\tau$, 
of the heavy neutrino.
Using this we can write down the Boltzmann equations for the evolution
of distribution functions
\begin{equation}
{\partial f}/{\partial t}- H p \, {\partial f}/{\partial p}=
C_{\mbox{\scriptsize{dec}}}[f].
\end{equation}
The decay terms are of 
the form \cite{kaiser}

\begin{eqnarray}
C_{\mbox{\scriptsize{dec}}}[f_{H}] & = &
- \frac{m_{H}^{2}}{\tau m_{0} E_{H}
p_{H}}
\int_{E_{\phi}^{-}}^{{E_{\phi}^{+}}}dE_{\phi}
\Lambda(f_{H},f_{L},f_{\phi}), \\
C_{\mbox{\scriptsize{dec}}}[f_{L}] & = &
\frac{g_{H}}{g_{L}} \frac{m_{H}^{2}}{\tau m_{0} E_{L} 
p_{L}}
\int_{E_{H}^{-}}^{{E_{H}^{+}}}dE_{H}
\Lambda(f_{H},f_{L},f_{\phi}), \\
C_{\mbox{\scriptsize{dec}}}[f_{\phi}] & = &
\frac{g_{H}}{g_{\phi}}
\frac{m_{H}^{2}}{\tau m_{0} E_{\phi} 
p_{\phi}}
\int_{E_{H}^{-}}^{{E_{H}^{+}}}dE_{H}
\Lambda(f_{H},f_{L},f_{\phi}),
\end{eqnarray}
where
$\Lambda(f_{H},f_{L},f_{\phi}) = f_{H}(1-f_{L})(1+f_{\phi})-
f_{L}f_{\phi}(1-f_{H})$ and
$m_{0}^{2} = m_{H}^{2}-2(m_{\phi}^{2}+m_{L}^{2})+
(m_{\phi}^{2}-m_{L}^{2})^{2}/m_{H}^{2}$.
$\tau$ is the lifetime of the heavy neutrino and $g$ is the
statistical weight of a given particle. We use $g_{H} = g_{L} = 2$ and
$g_{\phi} = 1$, corresponding to $\phi = \overline{\phi}$ 
However, we could equally well have used $g_\phi=2$ without changing
the final results. Note that in all actual calculations we assume
$m_L = m_\phi = 0$ when solving the Boltzmann equation
(this does not influence the results at all). 
When calculating
transfer functions and CMBR power spectra we of course use the correct
value for $m_L$.
The integration limits are
\begin{eqnarray}
E_{H}^{\pm} (E_{i}) & = & \frac{m_{0}m_{H}}
{2m_{i}^{2}}\left[E_{i}\left(1+4\frac{m_{i}^2}{m_{0}^2}\right)^{1/2} 
\pm p_i \right], \\
E_{i}^{\pm} (E_{H}) & = & \frac{m_{0}}{2m_{H}}
\left[E_{H} \left(1+4\frac{m_{i}^2}{m_{0}^2}\right)^{1/2} 
\pm p_{H}\right],
\end{eqnarray}
where $i = L,\phi$.
We can vary the 
heavy neutrino mass without changing the final particle spectra if
we keep constant the ``relativity'' parameter of the decay, defined as
\cite{hannestad2}, 
$\alpha \equiv 1/9 \, ({m_H}/
{\text{MeV}})^{2} ({\tau_H}/{\text{s}})$.
This leaves us only one
free parameter that controls the decay kinematics and thereby the
shape of the particle distributions.
The dividing line between relativistic and non-relativistic decay is
$\alpha=1$ so that the decay is relativistic if $\alpha \lesssim 1$
and non-relativistic if $\alpha \gtrsim 1$. 
In Fig.~1 we show a realisation of this decay model for $\alpha=1$.
\begin{figure}[h]
\begin{center}
\epsfysize=7truecm\epsfbox{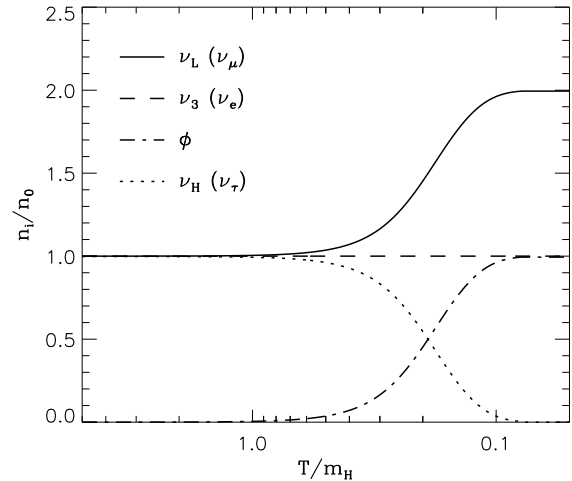}
\vspace{0truecm}
\end{center}
\baselineskip 17pt
\caption{The number density of different particle species as a function
of temperature for a model with $\alpha=1$. As explained we assume
$\nu_H=\nu_\tau$ and $\nu_L=\nu_\mu$. $\nu_e$ is assumed to be massless.
$n_0$ is the number density of a standard massless decoupled neutrino
species.}
\label{fig1}
\end{figure}

\begin{figure}[h]
\begin{center}
\epsfysize=7truecm\epsfbox{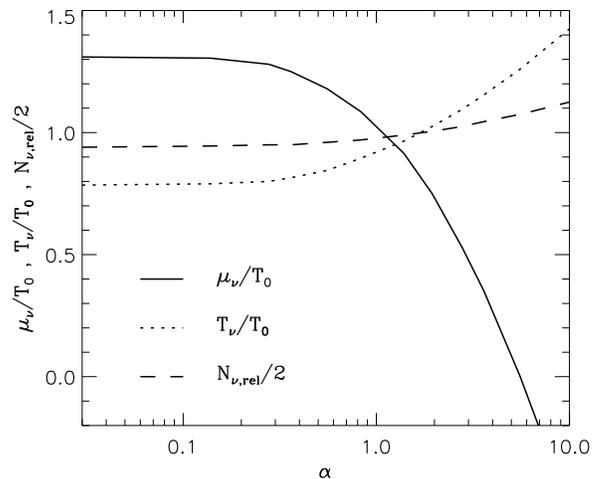}
\vspace{0truecm}
\end{center}
\baselineskip 17pt

\caption{The effective temperature, pseudo-chemical potential and
amount of relativistic energy density in massless neutrinos and scalar
particles. $T_0$ is the temperature of a standard massless decoupled 
neutrino species.}
\label{fig2}
\end{figure}

Now, to simplify matters we parametrise the final light neutrino
distribution in terms of an equilibrium Fermi-Dirac distribution with
an effective temperature, $T$, and a pseudo-chemical potential, $\mu$.
In Fig.~2 we show these two parameters in units of the temperature of
a standard massless neutrino species. Further, we show the amount of
relativistic energy density in massless
neutrinos (the energy density in the remaining third neutrino species)
and scalars in units of the
energy density of a standard neutrino species, 
$N_{\nu,\text{rel}}$. 
The energy density in light (HDM) neutrinos is not included here
because this neutrino becomes non-relativistic around matter-radiation
equality.
Notice that for relativistic decays 
$N_{\nu,\text{rel}}$ is smaller than two because the energy density in
scalars is less than in a standard massless neutrino species.
For relativistic decays all these parameters approach constant values
indicating that the decay proceeds in equilibrium, as expected.
For non-relativistic decays, however, we see that the effective 
temperature increases while the pseudo-chemical potential drops.
This reflects the fact that non-relativistic decays do not proceed
in thermodynamic equilibrium and the light neutrinos are born with 
energies much above thermal. Also, the amount of relativistic
energy density increases because the scalar particles, $\phi$, are
born with very high energies.
In fact, if the light neutrino mass approaches zero and the decay
is non-relativistic this scenario approaches the $\tau$CDM model
previously discussed in the literature \cite{DGT94,WGS95}.

To see how structure formation changes with different decay parameters
we have calculated the matter power spectrum in terms of the quantity
\cite{peacock},
$\Delta^2(k) \equiv \frac{k^3 P(k)}{2\pi^2}$,
by using the CMBFAST package modified to include a non-equilibrium
light neutrino species.
We have then compared these spectra with the linear power spectrum 
data compiled by Peacock and Dodds \cite{peacock}.
Fig.~3 shows the  power spectra for different models  
(note that all models shown in Figs.~3 and 4 have been calculated with 
$\Omega=1$, $\Omega_B=0.05$, $h_0=0.5$ and spectral index $n=1$).
It is seen that for relativistic decays 
and equal $\Omega_\nu$ our D$\nu$CHDM
power spectrum closely resembles that of the 2$\nu$CHDM scenario. This is
not too surprising because the light neutrino distribution is close to
thermal in this case. For non-relativistic decay, the difference is
pronounced. The power spectrum breaks away from that of the CDM model
at lower $k$-values if $\alpha \gg 1$ than if $\alpha \lesssim 1$ because
of the larger free-streaming length of the light neutrinos.
Even at large $k$ the power spectrum has less power if $\alpha$ is large.
This is because the amount of relativistic energy density is
significantly larger if the decay is non-relativistic.
The figure also shows that varying $\Omega_\nu$ influences
the power spectrum differently than changing $\alpha$, because varying
$\Omega_\nu$ changes the amount of CDM and the free-streaming length of
light neutrinos, whereas varying $\alpha$ changes the amount of
relativistic energy density and the free-streaming, but not the amount
of CDM.
We have also shown the limiting case of large $\alpha$ and low 
$\Omega_\nu$, where our model closely resembles the $\tau$CDM model.
As already shown in Ref.~\cite{DGT94} this model provides a very
good fit to the linear power spectrum.
\begin{figure}[h]
\begin{center}
\epsfysize=7truecm\epsfbox{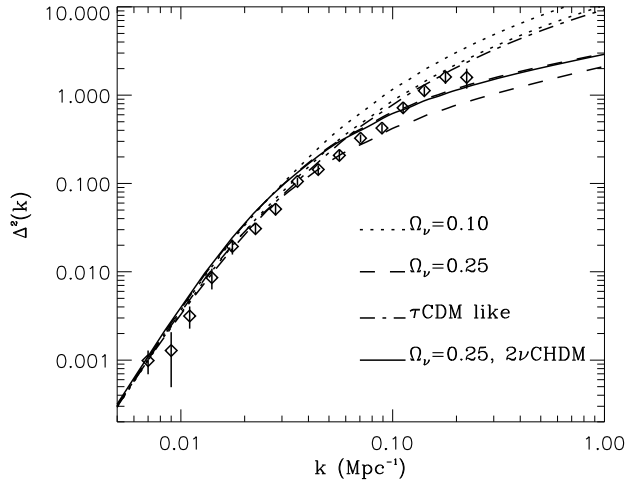}
\vspace{0truecm}
\end{center}
\baselineskip 17pt

\caption{Matter power spectra with HDM stemming from neutrino decay.
The two sets of lines correspond to $\alpha=0.1$ and 10, the upper 
curves being $\alpha=0.1$ and the lower $\alpha=10$.
For comparison we also show the 2$\nu$CHDM power
spectrum, as well as the $\tau$CDM-like power spectrum with 
$\Omega_\nu=0.01$, $\alpha=70$ and $h_0=0.45$.
The diamonds represent the compiled linear data of
Peacock and Dodds \protect\cite{peacock}.}

\label{fig3}
\end{figure}

Next, we also look at the power spectrum of CMBR anisotropies produced
by these scenarios. It has been known for some time that the CHDM
scenarios produce larger CMBR fluctuations than the CDM model and that
the peaks in the power spectrum are also shifted \cite{dodelson}.
We have calculated the power spectra for our series of models in terms
of $C_l$ coefficients,
$C_l \equiv \langle |a_{lm}|^2 \rangle$,
where the $a_{lm}$ coefficients are defined in terms of the temperature
fluctuations as
$T(\theta,\phi) = \sum_{lm} a_{lm} Y_{lm}(\theta,\phi)$.

Fig.~4 shows power spectra for several different models, illustrating
the differences between the CDM, 2$\nu$CHDM and D$\nu$CHDM models.
Exactly as with the matter transfer function we see that for relativistic
decays, the CMBR power spectrum for the D$\nu$CHDM model is very 
similar to that for the 2$\nu$CHDM scenario, 
indicating that the decay proceeds
roughly in equilibrium. For non-relativistic decays, however, the
power spectrum for the D$\nu$CHDM model is markedly different,
especially the amplitude of the first Doppler-peak is much larger. The
reason is that increasing $\alpha$ corresponds to increasing the 
relativistic energy density, meaning that recombination occurs much
closer to the radiation dominated epoch than in the standard model.
As explained for instance in Refs.~\cite{DGT94,DJ93}, the reason is
that in the matter dominated epoch photons travel through a constant
gravitational potential after they are emitted from the last scattering
surface. In the radiation dominated epoch this is not the case and
if the universe is still close to being radiation dominated at 
recombination the non-constant gravitational potential increases the
CMBR anisotropy, a feature known as the early ISW effect \cite{tegmark}.
Changing the HDM content by reasonable amounts in the different models 
only influences the CMBR power spectra slightly \cite{dodelson}.
\begin{figure}[h]
\begin{center}
\epsfysize=7truecm\epsfbox{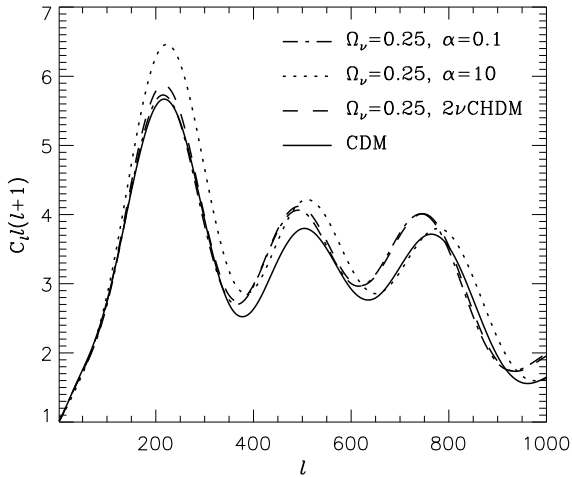}
\vspace{0truecm}
\end{center}
\baselineskip 17pt
\caption{CMBR power spectra for CDM, 2$\nu$CHDM and two different
D$\nu$CHDM models.
All the power spectra have been normalised to the quadrupole coefficient
$6C_2$.}

\label{fig4}
\end{figure}

In conclusion, we have seen that a natural way of producing a
$2\nu$CHDM like model 
is by decay of heavy neutrinos. If the decay proceeds 
relativistically this D$\nu$CHDM scenario closely resembles the standard 
2$\nu$CHDM scenario because particle distributions remain roughly
thermal. For non-relativistic decays there can be large differences,
both in the matter power spectra and in the CMBR fluctuation spectra,
compared with the 2$\nu$CHDM model. We have also shown that limiting
cases for our series of models is, in one end, the 
$\tau$CDM model \cite{DGT94},
where $\Omega_\nu \to 0$ and $N_{\nu,\text{rel}}$ is large 
(non-relativistic neutrino decays with very low light neutrino mass),
and in the other extreme the 2$\nu$CHDM model (relativistic decay with 
light neutrino mass equivalent to that in the 2$\nu$CHDM model)
\cite{PHKC95}.
A decaying $\tau$ neutrino can therefore
change structure formation on both small
and intermediate scales. Changing the time of matter-radiation equality
changes the amount of structure produced on horison-size scales at
this time, whereas changing the amount of neutrino hot dark matter
changes the amount of small scale structure.
There is some real hope that within the next few years it will become
possible to distinguish between these different models. Both because
new and improved galaxy surveys like the Sloan Digital Sky Survey
\cite{sloan}
are forthcoming and should greatly improve
our knowledge of the matter power spectrum and because the
next few years will hopefully see a leap in our understanding of the
cosmic background radiation because new missions are scheduled to fly
shortly, measuring the angular power spectrum to $l=1000$ and beyond
to great precision \cite{planck+map}. 
Using this new data it should be possible to
constrain structure formation models like the 
D$\nu$CHDM model described here very strongly. 

I wish to thank Jes Madsen for constructive criticism and discussions.
This work was supported by the Theoretical Astrophysics Center under
the Danish National Research Foundation.

\end{document}